\documentclass{aa}     

\usepackage{graphicx}
\usepackage{txfonts}
\usepackage{hyperref}

\usepackage{chemformula} 
\usepackage{lastpage}
\usepackage[svgnames]{xcolor}
\newcommand{\cmmo}{cm$^{-1}$}      
\newcommand{\pcmc}{cm$^{-3}$}      
\newcommand{\cmcs}{cm$^{3}/$s}     
\newcommand{\ie}{{\it i.e.}}
\newcommand{\eg}{{\it e.g.}}

\begin{document}

   \title{Collisional excitation of cyclopentadiene by helium}

   \subtitle{A complete set of rate coefficients and astrophysical applications}

   \author{S\'{a}ndor Demes \inst{1}\fnmsep\thanks{Current affiliation:
                             HUN-REN Institute for Nuclear Research,
                             Bem square 18/c, H-4026 Debrecen, Hungary}
          \and
          Fran\c{c}ois Lique\inst{1}
          \and
          Marcelino Ag\'{u}ndez\inst{2}
          \and
          Jos\'{e} Cernicharo\inst{2}
          }

   \institute{Univ Rennes, CNRS, IPR (Institut de Physique de Rennes) - UMR 6251,
              F-35000 Rennes, France\\
              \email{demes.sandor@atomki.hu; francois.lique@univ-rennes.fr}
         \and
              Instituto de Física Fundamental (IFF-CSIC), Dept. de Astrofísica
              Molecular, E-28006 Madrid, Spain\\
              \email{jose.cernicharo@csic.es}
             }

   \date{Received \@MONTH \@DD, 2026; accepted \@MONTH \@DD, \@YYYY}


  \abstract
   {Complex organic molecules, including large cyclic species, are prevalent in interstellar space and play a key role in various astrochemical processes. Cyclopentadiene (c-\ch{C5H6}) is a five-membered cyclic hydrocarbon recently detected in TMC-1 and some other interstellar molecular clouds. While accurate spectroscopic data were available, collisional rate coefficients for its rotational transitions were missing so far, introducing a potential limitation in the interpretation of the observations.}
   {This study aims to provide a comprehensive set of state-to-state thermal rate coefficients for the rotational excitation of c-\ch{C5H6} due to collisions with helium, crucial for non-local thermodynamic equilibrium (non-LTE) radiative transfer modelling in astrophysical environments, and to examine how far the molecule is from thermalisation under the physical conditions of cold molecular clouds.}
   {The research employed accurate quantum scattering calculations using the close-coupling (CC) and coupled states (CS) methods, based on a highly-correlated three-dimensional potential energy surface for the [c-\ch{C5H6 - He}] collisional complex. Calculations were performed for a wide range of rotational states (from $j=0$ to $j \leq 25$) and kinetic temperatures (from $10$ to $50$~K).}
   {We calculated a complete set of thermal rate coefficients for both {\it ortho} and {\it para} nuclear spin configurations of c-\ch{C5H6}. Radiative transfer simulations demonstrated that most rotational levels of cyclopentadiene are fully thermalized under typical cold cloud conditions and exhibit minor non-LTE effects. Nevertheless, this study is the first to utilise accurate state-to-state rate coefficients for radiative transfer simulation of a large, five-membered cyclic species detected in space.
   This allows to draw some general conclusions and paves the way for future collisional excitation studies of complex astromolecules that will enable a more precise interpretation of upcoming detections.}
  {}

   \maketitle
%
\section{\label{sec:Intro} Introduction}
Complex organic molecules (COMs), which form single or multiple rings (\ie~polycyclic aromatic hydrocarbons -- PAHs), are widespread in space. Early predictions have been made by \cite{chiar2013}, showing that collisions of stardust grains in shock waves might be a relevant source of interstellar PAHs. A series of large cycles were also detected recently {\it in situ} in the samples of the Ryugu asteroid \citep{Zeichner_2023} collected by the Hayabusa2 spacecraft. These complex species are known to be the main building blocks of organic chemistry on Earth, so their investigation in interstellar environments are essential to understand the formation and origin of biological molecules. While the astronomical detection of these complex species is difficult, the presence of simple five- and six-membered hydrocarbon rings as well as complex PAHs in molecular clouds of the interstellar medium (ISM) has been unambiguously proven now.

The first large aromatic carbon cycle detected in space was benzene \citep{Cernicharo2001}. Several years later a cyano-derivative of this molecule, the aromatic benzonitrile (c-\ch{C6H5CN}), was discovered by \cite{mcguire2018} in the Taurus Molecular Cloud (TMC-1). The first discovery of a pure hydrocarbon containing a five-membered cycle, c-\ch{C5H6} (cyclopentadiene), has been reported by \cite{cernicharo2021a}, along with other complex cycles c-\ch{C3HCCH} and c-\ch{C9H8} (indene). Individual rotational transitions of CN and CCH derivatives of c-\ch{C5H6} have been found by \citet{Cernicharo2021b}. Detection of these species by stacking techniques have been also reported \citep{mccarthy2020,kelvin2021}. Apart from the TMC-1 source, \cite{agundez2023} have found multiple large cyclic species in other cold environments, including cyclopentadiene along with its cyano-derivatives c-\ch{C5H5CN}.

\cite{He_2020} showed that c-\ch{C5H6} can be effectively formed in the gas phase from 1,3-butadiene (\ch{CH2CCHCH3}) in low-temperature environments typical to cold interstellar molecular clouds such as TMC-1. We note, however, that a systematic search for cyano derivatives of butadiene produced only negative results \citep{Agundez2025}, which introduce some doubts on the presence of 1,3-butadiene in TMC-1. Therefore, the chemistry of c-\ch{C5H6} is still rather poorly understood \citep{Agundez_2026}. Very recently, new chemical routes have been proposed by \citet{Jacovella_2026} to explain the interstellar formation of cyclopentadiene, which significantly improves the previous models, nevertheless it still reproduces only about $20\%$ of its observed abundance. As shown long before its first discovery in space, c-\ch{C5H6} could be also a key precursor of several PAHs including indene and naphthalene \citep{wang2006}. Cyclopentadiene therefore, along with its derivatives and other large cyclic and aromatic species play an important role in the chemistry of cold interstellar clouds. While pure rotational lines that identify particular PAHs have only been observed in cold molecular clouds so far (where the average temperature is $\sim10$~K), Diffuse Interstellar Bands (DIBs) are commonly associated with the near-UV to near-IR absorption signals from such complex species. However, their unambiguous identification remains an open question \cite{Lin_Yang_Li_2023, Salama_etal_1999}. In general, the rapidly growing number of observations of new complex molecular species significantly contributes to our understanding of the composition of interstellar matter and helps to build appropriate chemical models \citep{McGuire_2022}. 

To derive correct abundances for these large hydrocarbons, it is necessary to explore their spatial distribution \citep{Cernicharo2023} and to have a detailed molecular-level and line strengths description which are obtained from high-resolution microwave spectroscopy. Hence, accurate collisional data are also needed, as the molecular rotational levels are populated due to competing radiative and collisional processes in astronomical environments, where local thermodynamic conditions are usually not fulfilled (\ie~under non-LTE conditions, see for example \citealt{Cernicharo1987,Lique2006,Roueff2013} for details). Unfortunately, we have very sparse knowledge about the collisional excitation of large cyclic species, which is limited to some rate coefficients for cyanocyclopentadiene (c-\ch{C5H5CN}) calculated by \cite{Sogomonyan_2025}, benzene (\ch{C6H6}; no permanent dipole and, hence, no rotational spectrum \citealt{mandal2022}) and benzonitrile (c-\ch{C6H5CN}; only for rotational levels below 23 K, calculated from approximate scattering theories \citealt{Ben2024}).

The rotational spectra of cyclopentadiene (we simply refer to it as \ch{C5H6} hereinafter) is rather well-studied. \cite{Laurie1956} carried out the first high-resolution spectroscopy measurement, which uncovered as low as $0.416$~D dipole moment for this species. Later \cite{Damiani_1976} further studied the spectra of \ch{C5H6}, while the most accurate and commonly used spectroscopic properties have been measured by \cite{Bogey1988}. Cyclopentadiene is an asymmetric top with a $C_{2v}$ symmetry group. Therefore, its rotational levels are characterized by main ($j$) and projection ($k_a$ and $k_c$) quantum numbers that are usually denoted as $j_{k_a,k_c}$. These quantum numbers define the two possible nuclear spin configurations of c-\ch{C5H6}, in particular the {\it para}~($p$) one, when the sum $k_a + k_c$ is even, and {\it ortho}~($o$) in the opposite case (note that $k_a + k_c$ should be strictly equal to $j$ or $j+1$ and {\it ortho}-to-{\it para} transitions are forbidden).
For a more intuitive, graphical representation of the rotational structure  of cyclopentadiene, one can refer to Fig.~1 in our previous paper \citep{Demes_2024}, referred as \hypertarget{paper1}{Paper~I}  hereinafter.

The state-to-state collisional rate coefficients of c-\ch{C5H6}, which allow to perform non-LTE radiative transfer simulations, have not been studied before. Besides that, to the best of our knowledge, there are no radiative transfer calculations reported earlier for any five-membered cyclic hydrocarbons that is based on proper collisional rate coefficients. The present work aims to provide a full set of accurate rate coefficients for the rotational excitation of cyclopentadiene that is calculated for collisions with \ch{He} atom. Helium is known to be a suitable template for \ch{H2} in non-LTE simulation of complex molecular species \citep{Roueff2013}, which is currently not possible to treat as a projectile (along with a COM target) using reliable quantum scattering theories. Our calculations are carried out by means of the close coupling (CC) and coupled states (CS) quantum scattering theories and are based on an accurate 3-dimensional (3D) potential energy surface (PES), introduced in \hyperlink{paper1}{Paper~I}. In the present work, we have significantly extended the total energy range of our former study (from $60$ up to $500$~\cmmo), as well as the number of rotational states that are targeted (all levels are considered now up to $\sim100$~\cmmo , involving those with $j \leq 25$), which allows to derive state-to-state thermal rate coefficients for kinetic temperatures from $10$ to $50$~K (much above the typical temperatures in cold interstellar clouds). The new collisional data are used for non-LTE radiative transfer simulations, which are also demonstrated in this work. The paper is structured as follows: Section~\ref{sec:methods} provides the details of the scattering and radiative transfer model and methods, multiple subsections of Section~\ref{sec:Discuss} discuss the collisional cross sections and rate coefficients, Section~\ref{sec:radtrans} shows the results of radiative transfer simulations, while the final concluding remarks are given in Section~\ref{sec:concl}.

\section{\label{sec:methods} Model and methods}

In \hyperlink{paper1}{Paper~I}, we described the most important details about the interaction potential and the scattering methods used to treat the \ch{[C5H6 - He]} collisional system. We briefly summarise again the key aspects and provide the details about the extension of the model and methods, which allowed to calculate the rate coefficients that are presented in this work.

\subsection{\label{subsec:PES} PES for the \ch{C5H6 - He} interaction}
The dynamical calculations were performed using our original 3D rigid-rotor PES. It was calculated by the accurate CCSD(T)-F12b (\ie~the explicitly correlated coupled-cluster theory with singles and doubles and perturbative corrections for triple excitations) {\it ab initio} quantum chemical approach, using the augmented correlation-consistent polarised valence-triple-$\zeta$ (aug-cc-pVTZ) basis set. This level of theory ensures a good PES quality for a reasonable computational cost \citep{Demes_2020, Derbali_2023, Sunaga_2025}. The collisional complex is defined in a Jacobi coordinate system within the molecular frame representation (the origin is in the centre of mass of the rigid \ch{C5H6}, while the $\{R,\theta,\phi\}$ coordinates define the relative position of \ch{He} around it). An analytical description of the PES has been derived then using the following formalism:
\begin{equation}
V(R, \theta, \phi) = \sum_{l}^{l_\mathrm{max}} \sum_{m}^{l} V_{lm}(R)\frac{Y_l^m(\theta, \phi) + (-1)^m  Y_l^{-m}(\theta, \phi)}{1+ \delta_{m,0}} .
\label{eq:vrtp}
\end{equation}
Here $V_{lm}(R)$ are the radial coefficients, $Y_l^m(\theta, \phi)$  are the normalized spherical harmonics, $\delta_{m,0}$ is the Kronecker $\delta$-function, while $l$ and $m$ define the degree and order of the spherical harmonics ($m \leq l$ and is required to be a multiple of 2 due to the $C_{2v}$ symmetry of the target. We employed a standard linear least square fit over 12 540 {\it ab initio} single-point energies, based on a total of 81 $V_{lm}(R)$ functions. The largest spherical harmonics order considered is $l = 16$ and $m = 16$. The quality of the analytical expansion is fully suitable for quantum scattering calculations, its root mean square percentage error is always below $1\%$. The global well of the PES is rather shallow, only about $-90$ cm$^{-1}$, defining a geometry that is typical for a loosely bound complex. We cannot assess the accuracy of the interaction potential based on experiments, but similar PES calculations by \cite{Faure_2019} and \cite{Derbali_2023} show that the overall accuracy is expected to be on the wavenumber level. All other fine details about the PES, its features and the analytical expansion are described in \hyperlink{paper1}{Paper~I}.

\subsection{\label{subsec:ScattMeth} Molecular scattering calculations}

The analytical interaction potential for \ch{C5H6 - He} has been implemented successfully in the \texttt{MOLSCAT} scattering code \citep{Hutson_2019}, allowing to calculate accurate state-to-state rotational (de-)excitation cross sections. This was partially introduced in \hyperlink{paper1}{Paper~I}, which presented the results of accurate quantum scattering calculations. In this work, however, several crucial improvements have been done compared to the previous one, which is primarily related to astrophysical needs. First, the number of rotational levels has been drastically increased both for {\it o}/{\it p-} nuclear species of cyclopentadiene, from 22 up to more than 220. Consequently, all levels with rotational energies below $100$~\cmmo~($\sim141$~K) are covered now. It is worth mentioning that the lowest anharmonic vibrational frequency of \ch{C5H6} is about $520$~cm$^{-1}$ \citep{Alparone_2006}, so our rigid-rotor PES should be valid for all these rotational levels. Next, we have also significantly increased the total energy range, up to $500$~\cmmo~(in contrast with $\leq 60$~\cmmo~in \hyperlink{paper1}{Paper~I}), which allowed to derive a full set of thermal rate coefficients form $10$ to $50$~K. We used the exact close coupling (CC) quantum scattering method proposed by \citet{Arthurs_Dalgarno1960} to calculate the cross sections between states with internal energies below $25$~\cmmo (covering all potentially populated levels in cold molecular clouds).
Due to computational limits, the higher-lying levels (with internal energies from $25$ to $100$~\cmmo) have been considered through scattering calculations by the approximate and somewhat less accurate coupled states (CS) quantum scattering method \citep{McGuire_1974}, which neglects the Coriolis couplings between the particular rotational levels, and therefore the number of equations that needs to be solved can be significantly reduced. For all scattering calculations, the experimental rotational constants measured by \cite{Bogey1988} have been used: $A = 0.281$ \cmmo, $B = 0.274$ \cmmo~and $C = 0.142$ \cmmo. From Ray's formula $\kappa = (2B-A-C)/(A-C)$, one can derive an asymmetry parameter of $\kappa \simeq 0.9$ for cyclopentadiene, which indicates that it is close to an oblate symmetric top ($\kappa=1$). A detailed list of all {\it o}/{\it p-}\ch{C5H6} levels, which have been covered in the present work (all states with $j \leq 18$ as well as some of those up to $j = 25$) are provided in the electronic supplementary material. A reduced mass of $\mu = 3.774076$~amu has been calculated and used for the [\ch{C5H6 - He}] complex.

For practical reasons, we carefully made some truncations on the rotational basis set ($j_\mathrm{max}$), the maximal total angular momentum ($J_\mathrm{tot}$) as well as on the propagator parameters. The corresponding boundaries for these parameters were found based on systematic convergence test calculations, with a maximal relative error criteria of $\pm 1.0\%$ in the inelastic cross section for $j_\mathrm{max}$. This allowed to find the optimal rotational basis ranging from $j_\mathrm{max} = 24$ at the lowest collision energies, up to $j_\mathrm{max} = 35$ in the high-energy regime. In the case of $J_\mathrm{tot}$, we used the built-in automatic convergence-test module of \texttt{MOLSCAT} with \texttt{DTOL}=0.3 and \texttt{OTOL}=0.005 parameters (consequently, the largest allowed relative error for elastic partial cross sections is 0.3~\AA, and it is as low as 0.005~\AA for the inelastic ones).
Under such criteria, $J_\mathrm{tot}$ reached as high values as $80$ in the high-energy end. The propagator parameters (\texttt{RMIN}, \texttt{RMAX}, and \texttt{RMID}, see \cite{Hutson_2019} for details) have been selected based on a criteria that allows a maximum of $\pm 0.3 \%$ deviation in the inelastic cross sections. Due to the large rotational basis, there are many high-lying rotational levels that do not contribute significantly to the scattering cross sections of the target states. In order to neglect these, we applied an upper limit $E_ \mathrm{max}$ on the rotational basis, which is collision-energy-dependent and is also found through convergence tests with a criteria of $\pm 0.3 \%$ max. In the low-energy regime, where resonance structures are typical, we used the same fine collision energy grid that was used in \hyperlink{paper1}{Paper~I}, with a maximum of $0.1$~cm$^{-1}$ step size. The step size has been gradually increased at total energies above $\sim 75$~cm$^{-1}$ in the case of CC and above $\sim 150$~cm$^{-1}$ in the case of CS calculations (note that CS calculations covers a larger number of states, with the highest internal energy of $\sim 100$~cm$^{-1}$). For calculations from $200$~cm$^{-1}$ to $500$~cm$^{-1}$, only a few energies have been considered, since all cross sections show a monotonic behaviour in this regime.

We estimate the accuracy of the collisional data reported in this work is better than $\sim 10\%$ in the case of the numerically exact CC calculations and better than $25-30\%$ in the case, when the approximate CS method is used. Nonetheless, for the majority of transitions the accuracy is typically as good as $1-2\%$ (CC) and $10-20\%$ (CS), estimated as an accumulated uncertainty from all sources of errors, including the {\it ab initio} calculations, the analytical fit of the PES and the effect due to the optimizations and truncations in the scattering calculations.

\section{\label{sec:Discuss} Results and Discussion}

\subsection{\label{subsec:Crosssec} Cross sections}

\begin{figure*}[ht]
\centering
\includegraphics[width=0.49\linewidth]{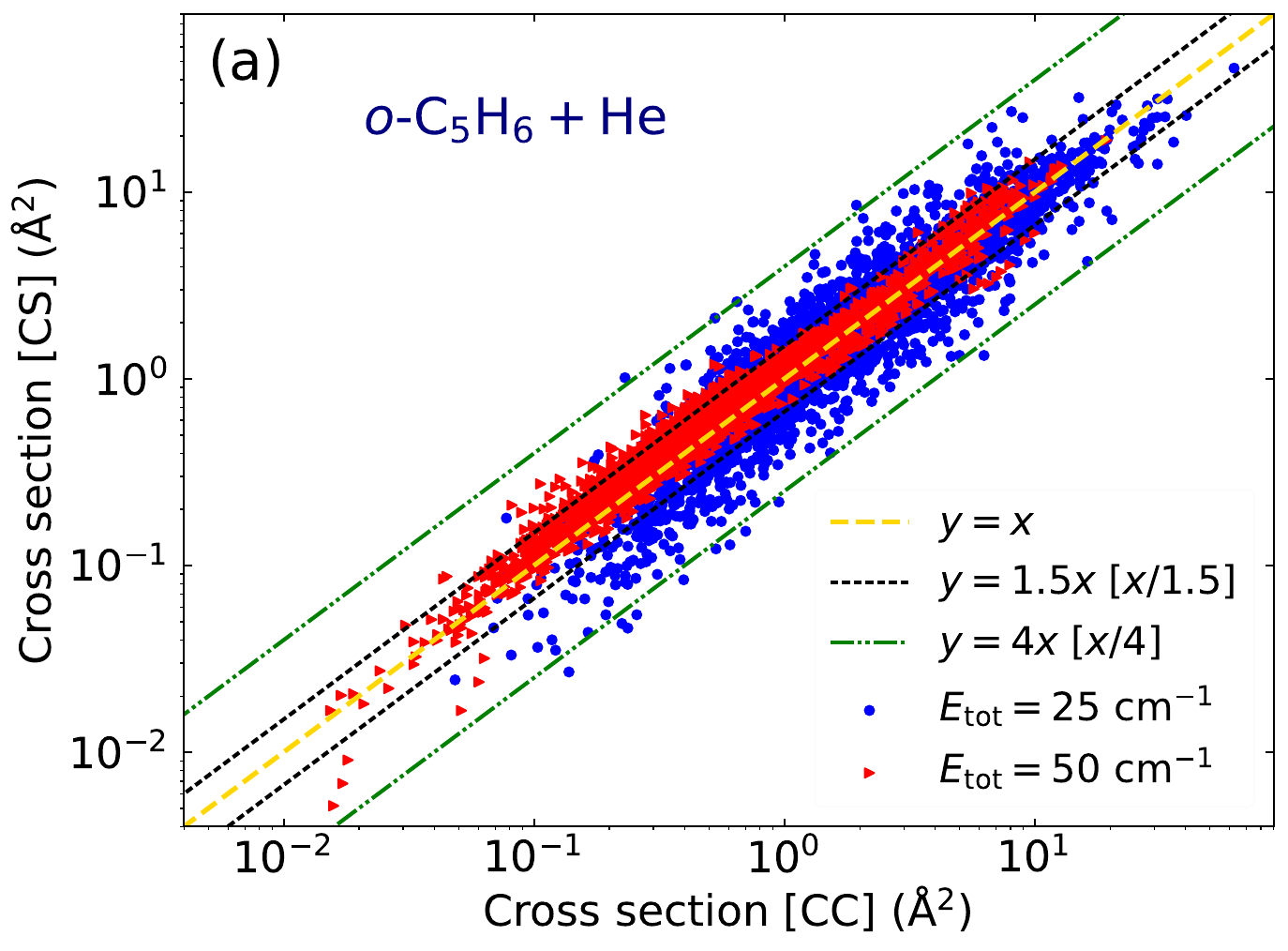}
\includegraphics[width=0.49\linewidth]{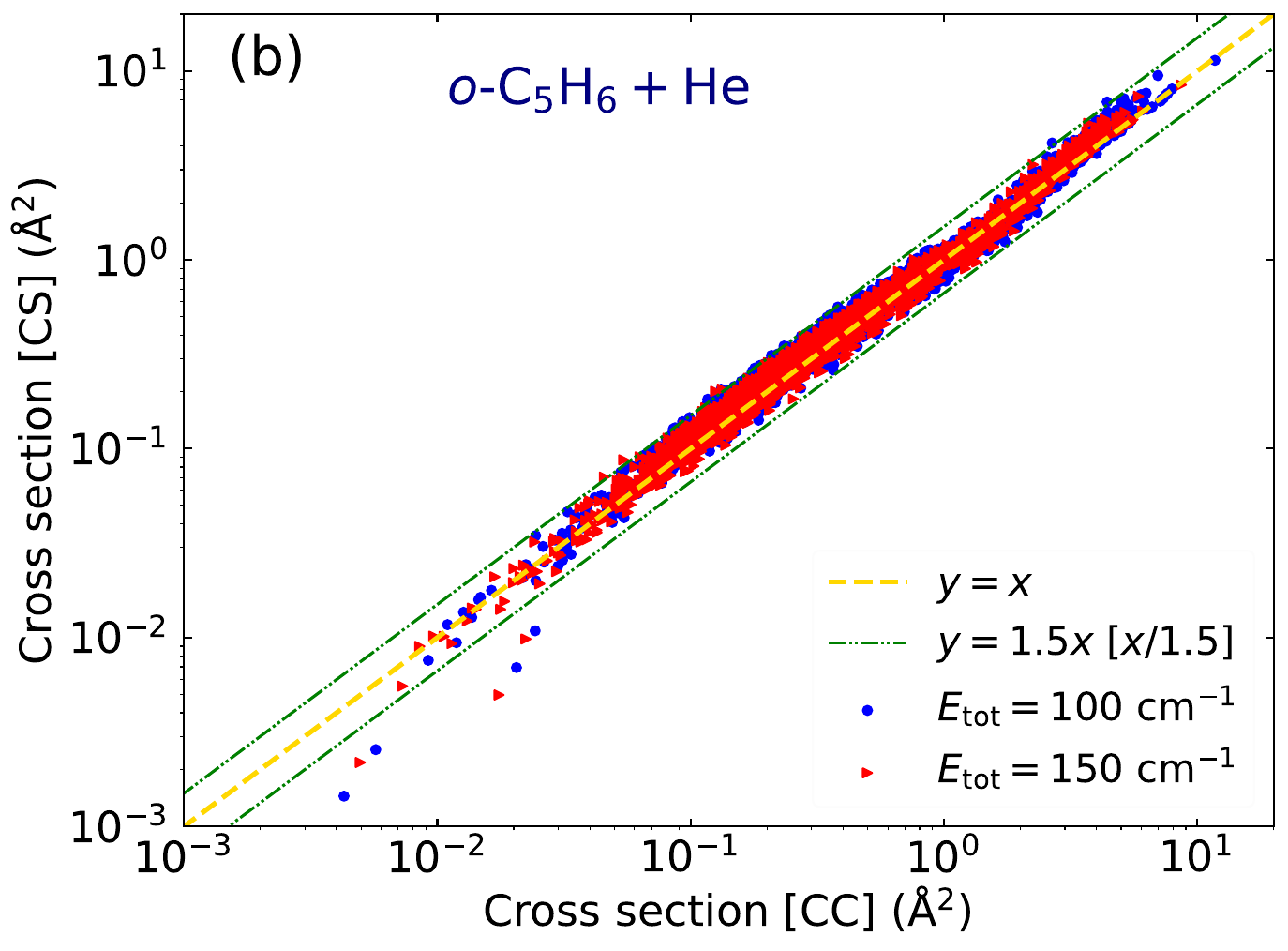}
\caption{State-to-state rotational (de)-excitation cross sections of  {\it ortho}-\ch{C5H6}  due to collisions with He, as calculated from the exact close coupling (CC) {\it versus} the approximate coupled states (CS) method. The left panel (a) compares the corresponding data in the low-energy scattering regime (total energies of $25$~cm$^{-1}$ and $50$~cm$^{-1}$), while the right panel (b) illustrate the results at higher energies, at $100$~cm$^{-1}$ and $150$~cm$^{-1}$.}
\label{fig:XS_ortho}
\end{figure*}

In \hyperlink{paper1}{Paper~I}, we have analysed the rotational (de-)excitation cross sections for low-energy \ch{C5H6 - He} collisions. We identified the most important propensity rules that mostly favour $\Delta k_c = 0$ transitions. We have also shown that notable deviations can be found in the cross sections when the approximate CS theory is used rather than the exact CC method in the low-energy regime. As the number of states have been significantly increased in the current work, we performed a more detailed state-to-state analysis of the cross sections at various energies, which is depicted in Fig.~\ref{fig:XS_ortho}. Its two panels show the low-energy and high-energy scattering regimes separately. In particular, Fig.~\ref{fig:XS_ortho}.a compares {\it ortho}-\ch{C5H6 - He} cross sections at $25$~cm$^{-1}$ and $50$~cm$^{-1}$ total energies, as calculated from the CC ($x$-axis) and CS theories ($y$-axis). The right panel (Fig.~\ref{fig:XS_ortho}.b) then illustrates the corresponding cross sections at higher energies ($100$~cm$^{-1}$ and $150$~cm$^{-1}$), outside the PES well and resonance regions.

As one can see, there are rather significant deviations between the cross sections calculated from the two quantum theories at $25$~cm$^{-1}$ and $50$~cm$^{-1}$. These deviations are usually within a factor of $1.5-2$, but for some transitions, differences up to a factor of $\sim4$ can also be observed. The mean absolute error (MAE) can be defined as:
\begin{equation}
\mathrm{MAE} = \frac{1}{N}\sum_{i=1}^{N} \frac{|\sigma_i^\mathrm{CC} - \sigma_i^\mathrm{CS}|}{\sigma_i^\mathrm{CC}} \times 100\% ,
\label{eq:mae}
\end{equation}
where $N$ is the total number of inelastic transitions at a specific collision energy, $\sigma_i^\mathrm{CC}$ and $\sigma_i^\mathrm{CS}$ are the state-to-state cross sections calculated from the CC and CS methods, respectively. The MAE was found to be $30.5\%$ and $19.9\%$ for {\it o}-\ch{C5H6} at $25$~cm$^{-1}$ and $50$~cm$^{-1}$, respectively. This trend is very similar in the case of {\it p}-\ch{C5H6}, with MAE values of $28.9\%$ and $21.1\%$, at the same energies respectively. The high-magnitude cross sections usually show somewhat better agreement, while there are no systematic trends, {\it i.e.} both over- and underestimations can be equally found. The overall picture is very similar for both nuclear species of \ch{C5H6}. It is also worth highlighting that the deviations are significantly smaller at $50$~cm$^{-1}$ as compared to $25$~cm$^{-1}$, as also supported by the corresponding MAE statistics. This is most probably related to the stronger resonance structure at $25$~cm$^{-1}$, where the CS approximation is more likely to produce large errors or miss the resonance peaks (see Fig.~8 and the related discussion in \hyperlink{paper1}{Paper~I}).

On the contrary to the low-energy-regime, we found a very good overall agreement between the cross sections calculated from the CC and CS quantum theories at higher energies, as it can be seen in the right (b) panel of Fig.~\ref{fig:XS_ortho}. The MAE value for {\it o}-\ch{C5H6} presented in this figure is $12.2\%$ at $100$~cm$^{-1}$, while it is even lower at $150$~cm$^{-1}$, about $10.0\%$ only. For {\it p}-\ch{C5H6}, these statistics are very similar again, with $12.6\%$ mean error at $100$~cm$^{-1}$, and $10.5\%$ at $150$~cm$^{-1}$ total energies. One can see a general trend that the discrepancies between the cross sections from the CC and CS methods tend to disappear at higher collision energies, especially at those larger than the global well depth ($\gtrsim 90$~cm$^{-1}$). We do not see any transitions here, where the relative difference between CS and CC cross sections is larger than a factor of 1.5. These findings justify the applicability of the approximate CS method for calculating rate coefficients at higher temperatures, where the dominant contributions are coming from the middle- and high-energy parts of the cross sections, as defined by the specific Boltzmann distribution.

\subsection{\label{subsec:Ratecoeff} Rate coefficients}

Once the deviation of the cross sections calculated from the CS and CC methods have been assessed, we also derived the corresponding thermal rate coefficients. These collisional data are essential for proper radiative transfer models. Rate coefficients are calculated up to a kinetic temperature of $50$~K by the usual integration over a Maxwell-Boltzmann distribution of relative energies as follows:

\begin{equation}
    k_{\mathrm{i} \rightarrow \mathrm{f}}(T) = \left(\frac{8}{\pi\mu k_\mathrm{B}^3 T^3}\right)^\frac{1}{2} \int_{0}^{\infty} E_\mathrm{kin} \,  e^{-\frac{E_\mathrm{kin}}{k_\mathrm{B}T}} \,  \sigma_{\mathrm{i} \rightarrow \mathrm{f}}(E_\mathrm{kin}) \,  \mathrm{d}E_\mathrm{kin} ,
    \label{eq:rates}
\end{equation}
where $E_\mathrm{kin}$ is the kinetic or collision energy, $\sigma_{\mathrm{i} \rightarrow \mathrm{f}}(E_\mathrm{kin})$ is the cross section that describes the transition efficiency from initial state (i) to final state (f), $\mu$ is the reduced mass of the [\ch{C5H6 - He}] collisional complex and $k_\mathrm{B}$ is the Boltzmann constant. All constituents of Eq.~\ref{eq:rates} are defined in atomic units.

\begin{figure}[ht]
\centering
\includegraphics[width=0.99\linewidth]{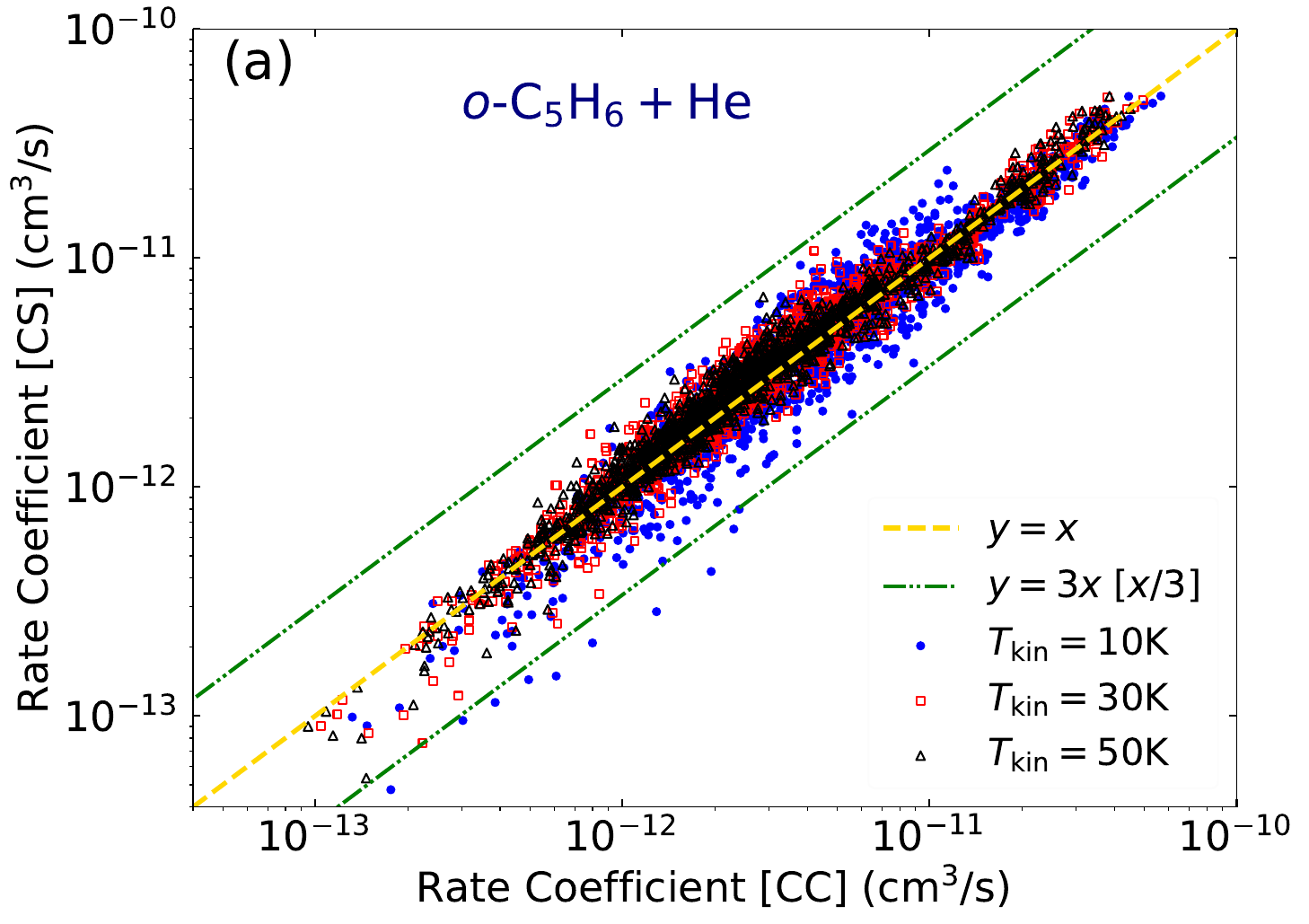}
\includegraphics[width=0.99\linewidth]{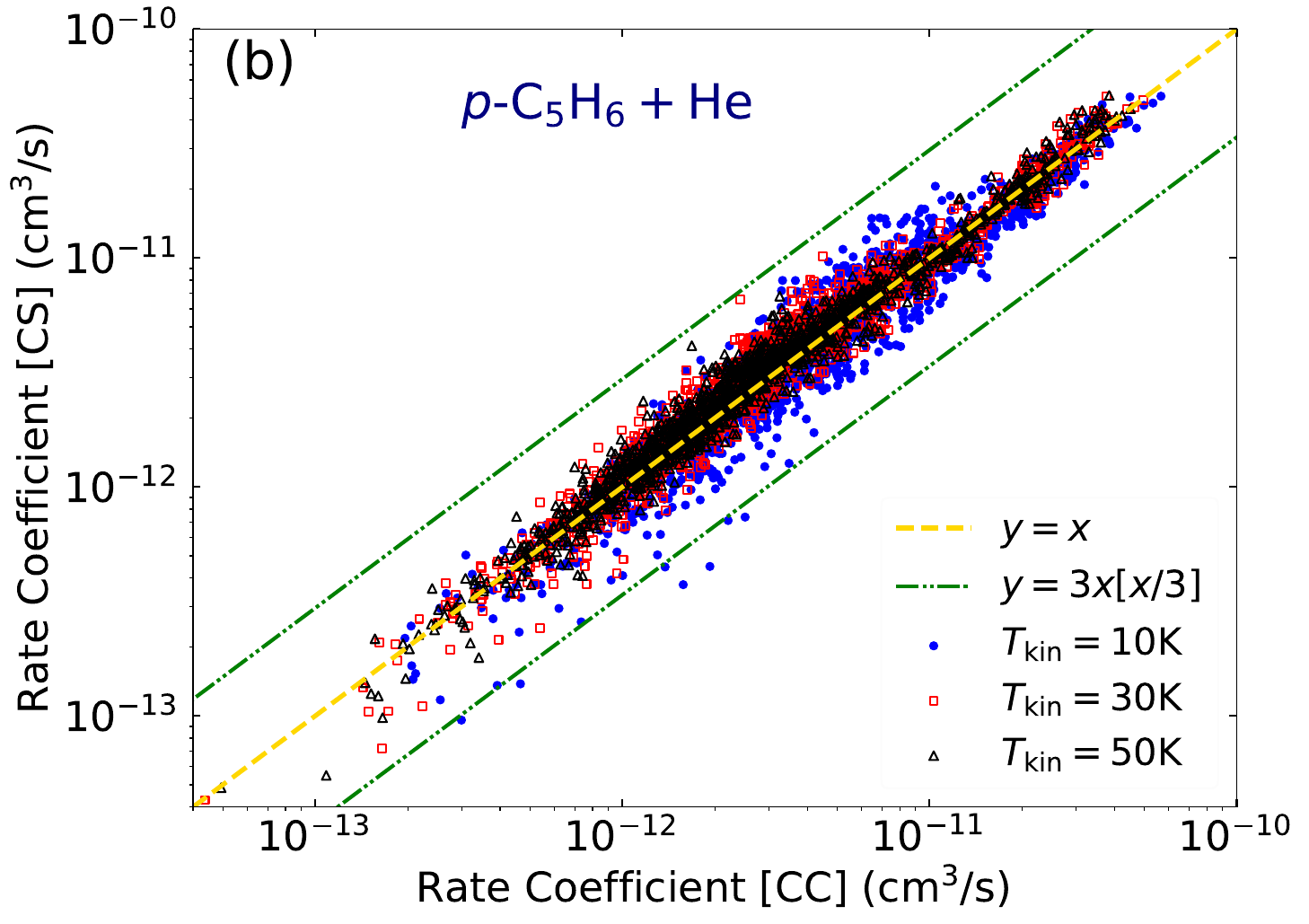}
\caption{Comparison of the state-to-state thermal rate coefficients for the rotational excitation of {\it ortho}-\ch{C5H6} (a) and {\it para-}\ch{C5H6} (b) due to collisions with He, as calculated from the CC and CS methods. Kinetic temperatures of $10, 30$ and $50$~K are depicted.}
\label{fig:RA_CCvsCS}
\end{figure}

\begin{figure}[ht]
\centering
\includegraphics[width=0.99\linewidth]{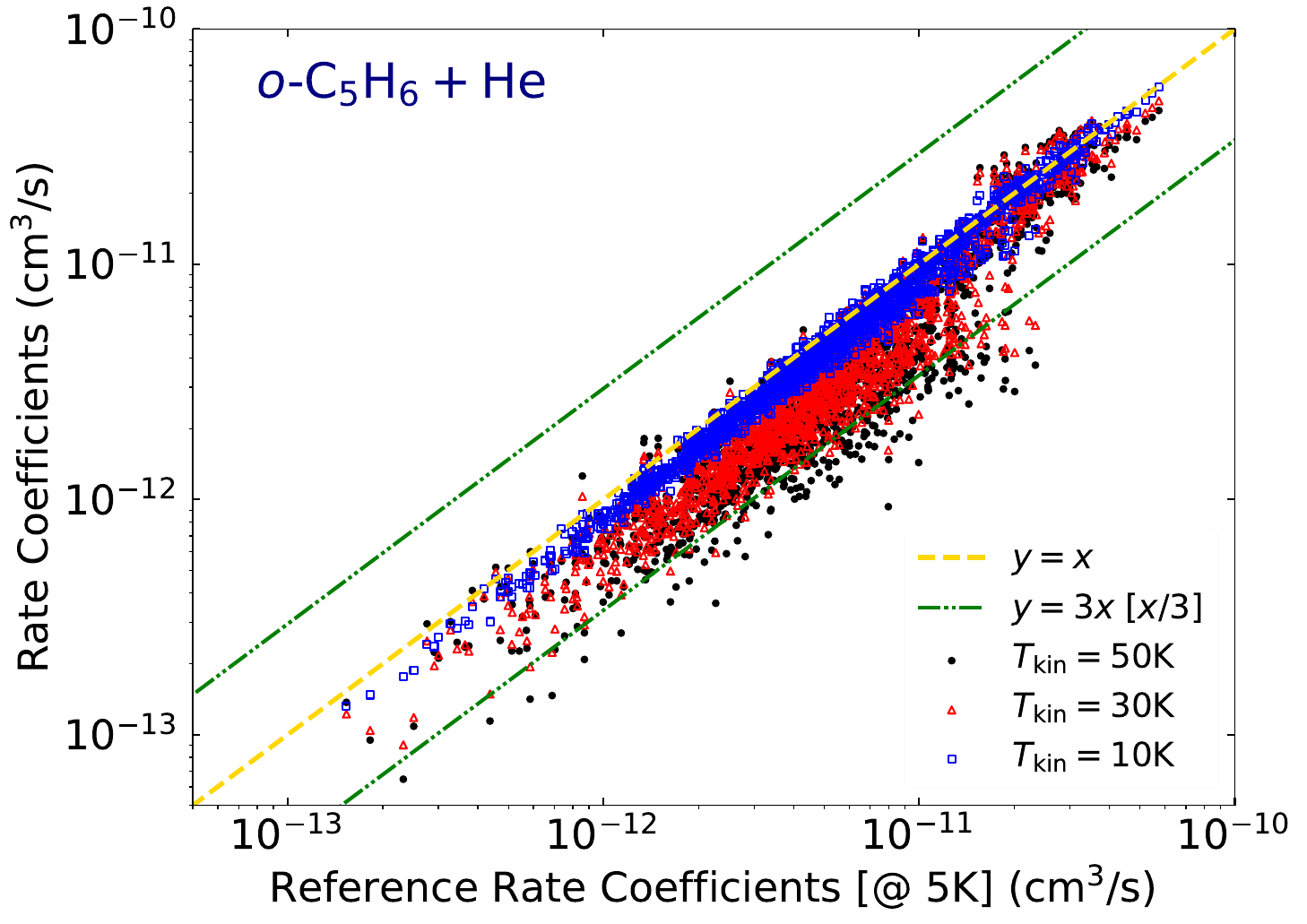}
\caption{Temperature dependence of the state-to-state rotational de-excitation rate coefficients for the {\it ortho}-\ch{C5H6 +He} collision at kinetic temperatures from $10$ to $50$~K with respect to the data derived at $T_{\mathrm{kin}} = 5$~K.}
\label{fig:RA_Tdep}
\end{figure}

\begin{figure*}[ht]
\centering
\includegraphics[width=0.48\linewidth]{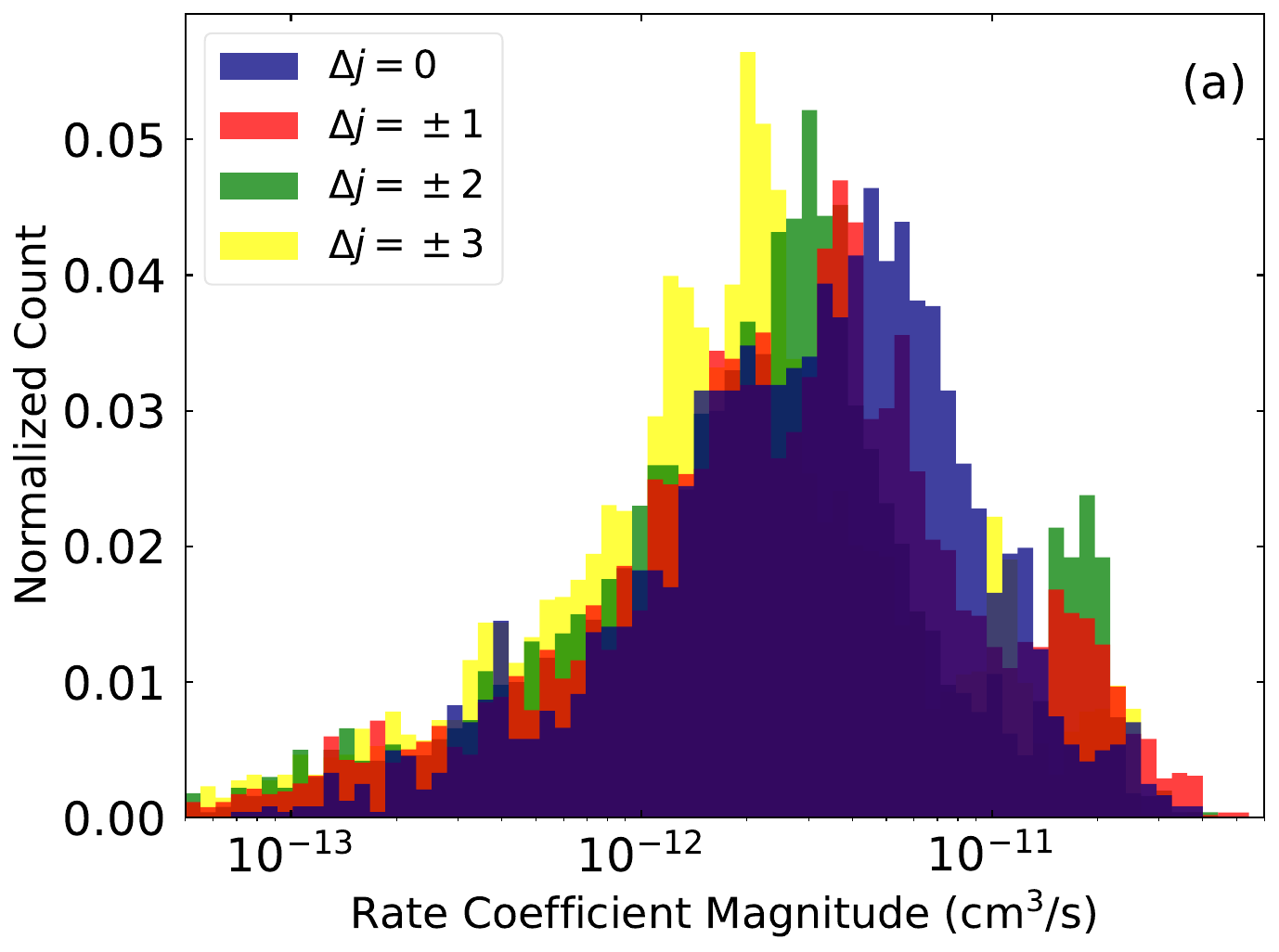}
\includegraphics[width=0.48\linewidth]{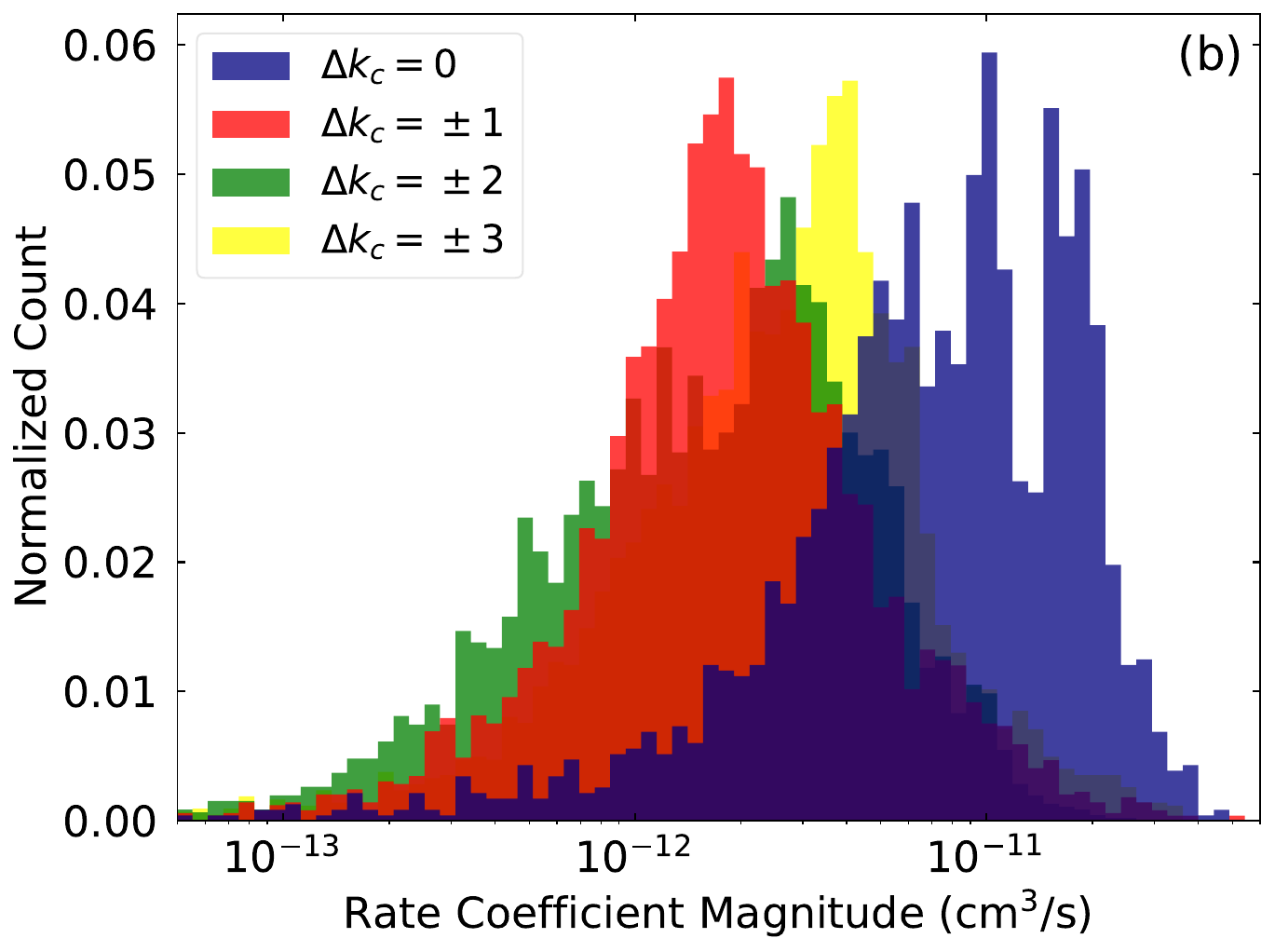}
\caption{Normalized distribution of all calculated rotational de-excitation rate coefficients for \ch{C5H6} with respect to the variation of the main ($\Delta j$, panel a) and projection ($\Delta k_c$, panel b) quantum numbers during the transitions at $10$~K.}
\label{fig:distro}
\end{figure*}

Fig.~\ref{fig:RA_CCvsCS} describes the comparison of the state-to-state collisional rate coefficients for cyclopentadiene at $10, 30$ and $50$~K kinetic temperatures, as derived from the CC and CS cross sectional data. The upper panel (a) shows the results for {\it ortho}-\ch{C5H6}, while the bottom panel (b) is for the {\it para} nuclear spin species. One can see that there is a relatively strong correlation between the rate coefficients calculated from the two quantum methods, even at the lowest temperature of $10$~K (typical for cold clouds, like TMC-1). The differences in the two datasets only barely exceeds a factor of 3, and are usually within a factor of $1.5$. The correlation is systematically stronger as the temperature increases, with MAE of about $22.4\%$ for both {\it o/p}-\ch{C5H6} at $10$~K, $17.5\%$ for {\it o}-\ch{C5H6} and $18.2\%$ for {\it p}-\ch{C5H6} at $30$~K, and as low as $15.3\%$ and $16.2\%$ at $50$~K for {\it o/p}-\ch{C5H6}, respectively. The relative differences are significantly smaller between the CC and CS rate coefficients in the case of the high-magnitude transitions ($>10^{-11}$~\cmcs), which are typically the most dominant ones in radiative transfer simulations. Deviations in both directions are present, so there is no systematic, correlated scaling between the CC and CS results. However, at $50$~K, the results from CS tend to often overestimate the more exact CC results. It is also worth noting that the magnitude of the rate coefficients for \ch{C5H6} is usually lower in contrast to those that are typical for smaller polyatomic species, and do not exceed even the value of $(5-6) \times 10^{-11}$~\cmcs. This may be due to its high density of states that reduces the probability of state-to-state transitions (the flux of the energy transferred due to collision is shared along many competing channels).
We examined the available data for other complex species in collisions with He, and we found very similar magnitudes that do not exceed $10^{-10}$ \cmcs. This is valid, among others, in the case of \ch{CH3CHCH2O} \citep{Dzenis_2022}, \ch{HCOOCH3} \citep{Faure_2014}, the cyclic c-\ch{C3H2} \citep{Khalifa2019} and also the data reported recently for the chemically similar cyanocyclopentadiene (c-\ch{C5H5CN}) by \cite{Sogomonyan_2025}.

\begin{figure*}[ht]
\centering
\includegraphics[width=0.99\linewidth]{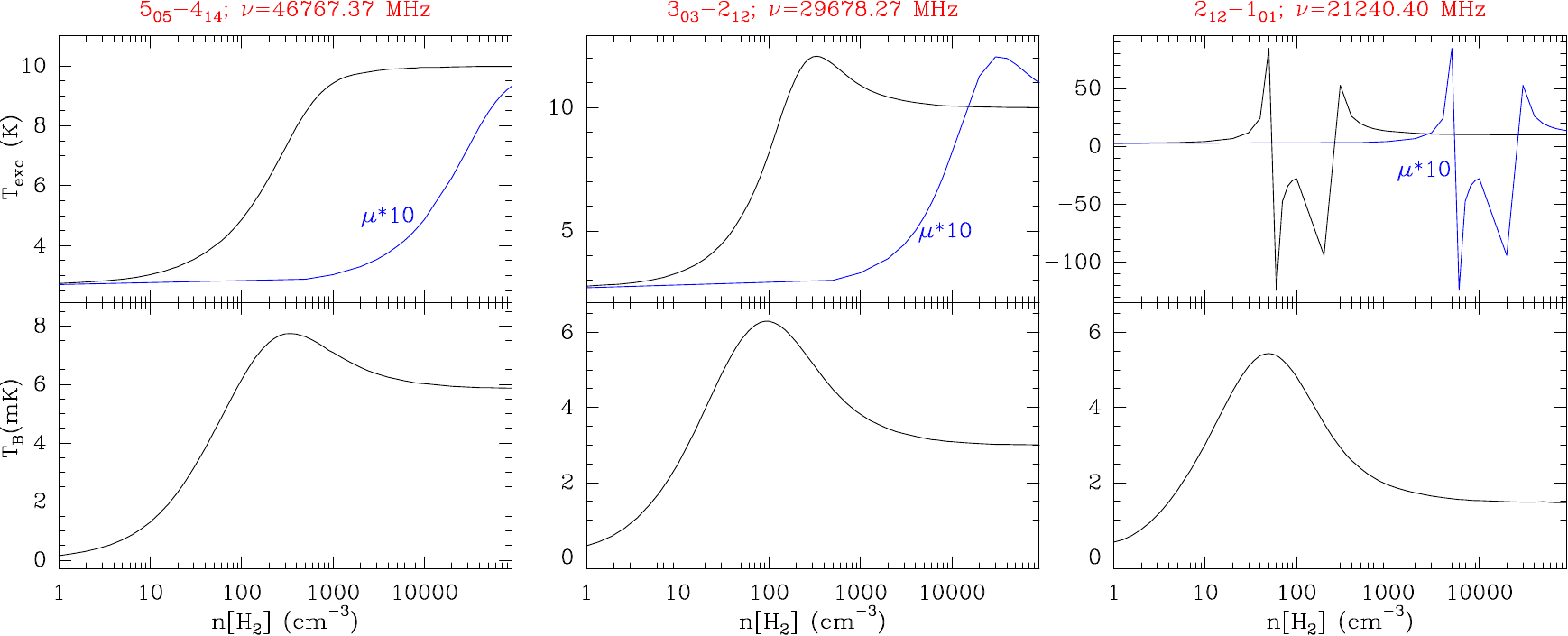}
\caption{Calculated excitation and brightness temperatures for three {\it ortho} transitions of cyclopentadiene.
We have assumed a kinetic temperature of 10 K, a column density of 10$^{13}$ cm$^{-2}$ and a line width
of 1.0 km\,s$^{-1}$. Under these conditions all lines are optically thin. Due to the low dipole moment
of the molecule, the rotational levels appear as nearly thermalised for densities between $400-2000$ cm$^{-3}$. The
2$_{12}$-1$_{01}$ transition shows weak maser effects for densities between 40 and 400 cm$^{-3}$.
For a molecule with a dipole moment 10 times larger (\ch{C5H5CN} for example) the values for the densities needed to obtain similar excitation
temperatures have to be multiplied by a factor 100 (blue curves in the $T_\mathrm{exc}$ panels).}
\label{fig:collisions}
\end{figure*}

We also examined the the temperature dependence of the rate coefficients, systematically comparing the relative change in the collisional data at $10, 30$ and $50$~K, with respect to low-temperature reference data calculated at $5$~K. The analysis shows a rather significant negative temperature dependence that is depicted in Fig.~\ref{fig:RA_Tdep}. As can be seen, the majority of the rate coefficients somewhat decrease even with a modest change from 5 to $10$~K, but these shifts are usually within a factor of about $1.5$ only. As the temperature further increases, much stronger temperature dependence is observed, typically with deviations up to a factor of $\sim 3 $ at $30$~K, and up to a factor of about $5$ at the highest temperature studied, $50$~K. These trends are probably connected to the dense rotational structure of cyclopentadiene. In simple words, at low temperatures ($\lesssim 10$~K), the collisional energy transfer can normally initiate transitions between a few low-lying states only, which might be efficiently populated at this specific temperature. This is especially valid for de-excitation transitions, where the initial state must be already populated prior to collision. As the temperature increases, the number of possibly populated rotational states drastically increases, and therefore the collision energy flux is shared between many transitions. Consequently, the collisional population/de-population of the individual rotational levels is less efficient in this higher temperature regime (note that this is most likely the reason of the lower magnitudes too). However, there is special trend in the case of the strongest, high-magnitude transitions (typically above $10^{-11}$~\cmcs). First, these are usually less sensible to the change in temperature, and second, they usually show a mixed positive and negative temperature dependence. A possible explanation of this can be related to the very prominent propensity rules that arise from the specific coupling schemes between the initial and final rotational quantum numbers and the constituents of the PES in the scattering formalism. In \hyperlink{paper1}{Paper~I}, the propensity rules that drive the interaction of \ch{C5H6} with helium have been discussed already, which uniquely highlight the most favourable rotational transitions. We showed that there is a strong favour of $k_c$-conserving transitions, {\it i.e.} those where the  $k_c$-projection quantum numbers of the initial and final states are preserved ($\Delta k_c = 0$). For these, the corresponding cross sections are usually by more than an order of magnitude larger compared to the less dominant ones. The derived propensity rules are general for the \ch{C5H6 - He} collision, apply both for the {\it ortho} and {\it para} species. From astrophysical perspective, it is important to point out that the most dominant transitions have relatively constant rate coefficients that are not strongly affected by temperature.

The large number of rotational transitions also allowed us to study the distribution of the rate coefficients by their magnitude. In order to find the global propensity trends for collisional excitation of \ch{C5H6}, we analysed the rate coefficients based on the relationship between the specific quantum numbers of the initial and final states of the particular transitions. The results are depicted in Fig.~\ref{fig:distro}, which shows the normalized distribution for $\Delta j$ ladders (panel a), and those correspond to various $\Delta k_c$-combinations (panel b) at $10$~K. For this, the combined set of all de-excitation rate coefficients for {\it o}/{\it p}-\ch{C5H6} has been used, which contains data for a total of 48841 transitions. A normalization is carried out by dividing the counts (for each bin) with the total number of transitions in the particular dataset for specific $\Delta j$ or $\Delta k_c$ combinations.

As can be seen in Fig.~\ref{fig:distro}.a, there are no general propensity trends observed with respect to $\Delta j$, that one would generally expect in accordance with the usual dipole selection rules that favour $\Delta j = 0, \pm 1$ transitions. The normalized distributions are very similar for all $\Delta j$ combinations, with peaks around $(3-6) \times 10^{-12}$~\cmcs~(lower for larger $\Delta j$ ladders, and the highest for $\Delta j = 0$). It is important to highlight two additional peaks around $2 \times 10^{-11}$~\cmcs, one for $\Delta j = \pm 1$ and an even stronger for the $\Delta j = \pm 2$ ladder. The state-to-state analysis shows that the vast majority of these transitions can also be categorized as a $k_c$-conserving one, which are therefore very favourable (only to mention some of them, these include the $6_{1,5} \rightarrow 5_{1,5}$, $6_{2,5} \rightarrow 5_{0,5}$ and $7_{2,6} \rightarrow 6_{0,6}$ transitions). In the case of the $k_a$ quantum number, the distributions are very similar to those presented for $\Delta j$, and there are no general propensities found that enhance any specific $\Delta k_a$ transitions. These trends are almost identical at $50$~K, with a slight change in the height and approximate position of the peaks of the corresponding distributions.

The picture is very different for $\Delta k_c$, for which very pronounced propensity trends are found. These strongly favour the $\Delta k_c = 0$ transitions, as it was already explored in \hyperlink{paper1}{Paper~I} partially. In Fig.~\ref{fig:distro}.b, one can see that the $k_c$-conserving transitions have globally by factors of $5-10$ times larger magnitudes in general, compared to those with $|\Delta k_c| > 0$, estimated based on the normalized area under the histograms and on the approximate position of the largest peaks. It is important noting that $\Delta k_c = \pm 3$ transitions are generally stronger than those of $\Delta k_c = \pm 1$ and $\pm 2$, which might be related to strong couplings between some specific terms of the interaction potential. A few surprisingly favourable transitions are also found with $\Delta k_c = \pm 5$ and large initial main rotational quantum number $(j=9-12)$, which are not presented in Fig.~\ref{fig:distro}. These strong transitions might be important from an astrophysical perspective, however, the origin of such propensity trends requires further investigation.

Despite cyclopentadiene has only been detected in cold clouds so far, the new set of collisional rate coefficients allows to cover a broader range of temperatures, at least up to $50$~K, which might be important for future possible detections in slightly warmer environments, including those typical to DIBs, and the external layers of circumstellar envelopes.

\section{\label{sec:radtrans} Radiative transfer results}
We carried out the radiative transfer calculations under the large velocity gradient (LVG) formalism developed by \citet{Goldreich_Kwan_1974} for molecular clouds. We have used the \texttt{MADEX} code \citep{Cernicharo_2012} to achieve these calculations using the rotational constants derived from a fit to all available laboratory data
\citep{Laurie1956,Scharpen1965,Flygare1970,Bogey1988,Bonah2025}. 
We implemented the newly calculated set of rate coefficients for the \ch{C5H6 - He} collisional system. They have been corrected for collisions with H$_2$ adopting a He/H abundance of 10\% and multiplying the rates by the ratio of the reduced masses $\mu_\mathrm{[\ch{C5H6} + \ch{H2}]}/\mu_\mathrm{[\ch{C5H6} + He]}$. 
The calculated excitation and brightness temperatures for three {\it ortho} transitions of \ch{C5H6} in the QUIJOTE frequency coverage
are shown in Fig. \ref{fig:collisions}.
We have assumed a kinetic temperature of 10 K, a column density of 10$^{13}$ cm$^{-2}$ and a line width
of 1.0 km\,s$^{-1}$. Under these conditions all lines are optically thin.
As expected from the low dipole moment of the molecule \citep[0.419\,D,][]{Scharpen1965},
the excitation temperature of practically all transitions with upper energy levels below 40 K are close to the kinetic
temperature for densities as low as $\sim$10$^3$ cm$^{-3}$. Hence, the LTE calculations performed by
\citet{Cernicharo2021b} to derive the column density of cyclopentadiene are fully justified. An interesting result from Fig. \ref{fig:collisions} is that the
emission peak for all strong transitions in the Q-band are found for a density
$\sim$ 200 \pcmc, when thermalisation is not yet reached. This is due to the limited number of rotational levels populated at low densities (low value for the rotational
partition function). Hence, if the molecule is formed in all layers of
molecular clouds it could be possible to have a significant contribution of the
low density external layers to the emerging line profile. In such a case, LVG calculations are
fully needed to interpret the data.

\begin{figure}[ht]
\centering
\includegraphics[width=0.99\linewidth]{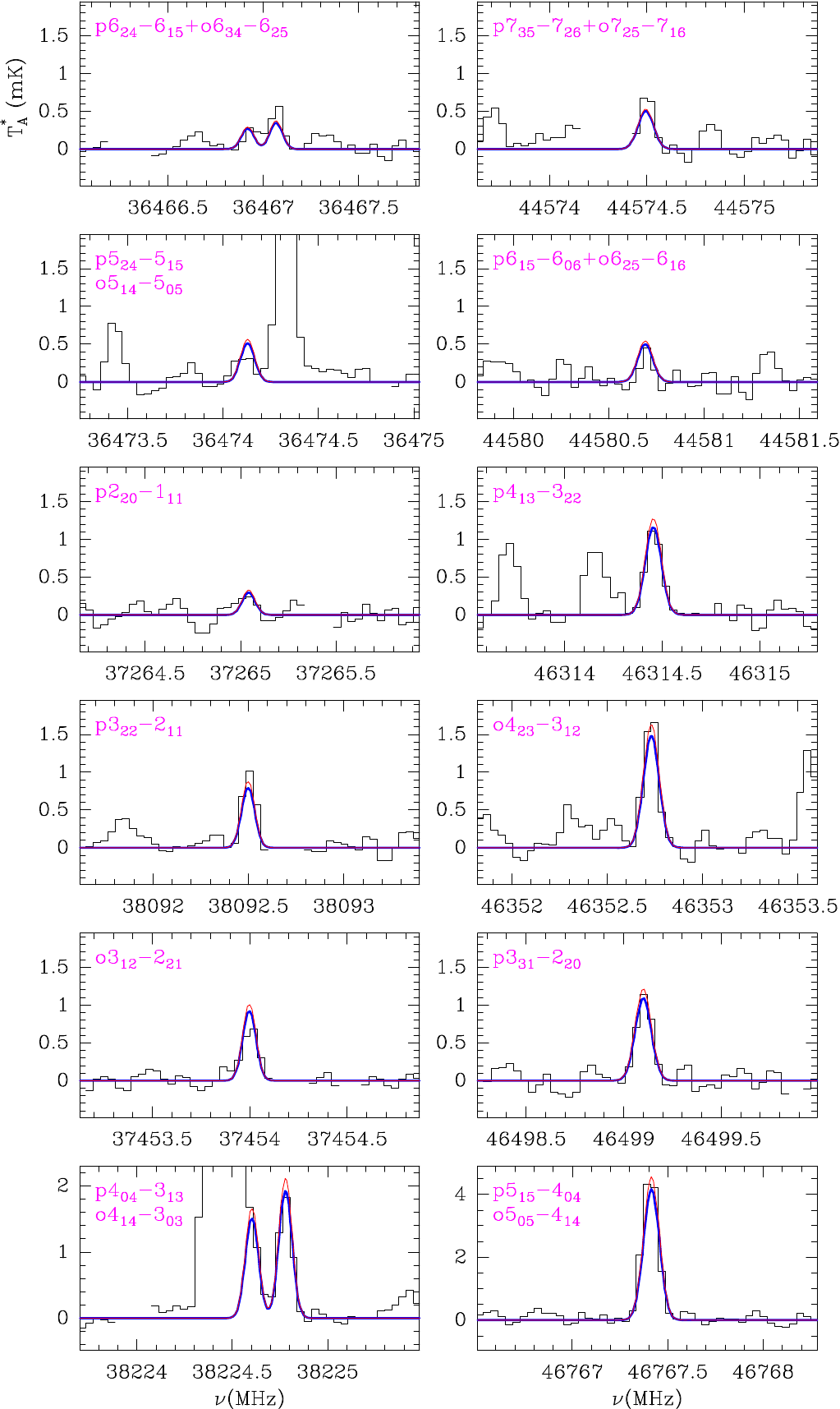}
\caption{All detected transitions of {\it o/p-}\ch{C5H6} observed by the last QUIJOTE line survey \citep{Cernicharo2021b,Cernicharo2026} The modelled non-LTE synthetic spectra are depicted in pink (assuming $n(\mathrm{H}_2)=2 \times 10^4$~\pcmc,~$T_\mathrm{kin}=9$~K and $\Delta v=0.6$~km/s). The synthetic spectra calculated from the LTE approach are presented in blue. Both calculated spectra, LTE and LVG, are nearly coincident.}
\label{fig:TMC-1_C5H6}
\end{figure}

In order to check which effect could be expected if the molecule had a large dipole moment, says 10 times larger (cyano cyclopentadiene for example), we have performed calculations for \ch{C5H6} adopting a hypothetical dipole moment of 4.2 D. The results are shown by the blue
curves in the excitation temperature panels of Fig. \ref{fig:collisions}. Under these conditions, densities around $\sim$10$^5$ \pcmc\, are needed to reach a near thermalisation level. Hence, it would be highly necessary to have collisional rate coefficients for the cyano-derivatives of cyclopentadiene, benzene, naphthalene and other PAHs detected in TMC-1.

In order to check observationally our results, we have used the last QUIJOTE\footnote{\textbf{Q}-band \textbf{U}ltrasensitive \textbf{I}nspection 
\textbf{J}ourney to the \textbf{O}bscure \textbf{T}MC-1 \textbf{E}nvironment}  
data
to compare LTE and LVG calculations for the c-\ch{C5H6} molecule. The QUIJOTE data correspond
to a spectral line survey of TMC-1 in the Q-band carried out with the Yebes 40m radio telescope  at the position $\alpha_{J2000}$= 4$^h$41$^m$41.9$^s$ and $\delta_{J2000}$=+25$^o$ 41$'$ 27.0$''$ (which corresponds to 
the classical cyanopolyyne peak in TMC-1). The QUIJOTE data have been
described by \citet{cernicharo2021a} and more recently by \citet{Cernicharo2026}. A detailed description of the system is given 
by \citet{Tercero2021}. Briefly, all observations were performed 
using frequency-switching observing mode with a frequency throw of 8 (736.6 hours) and 
10 MHz (772.6 hours). Hence, the total observing time on source is 1509.2 hours. The
measured sensitivity varies between 0.08 mK at 31 GHz and 0.2
mK at 49.5 GHz. 

The data of the QUIJOTE line survey presented
here were gathered in several observing runs between November 2019 and July 
2024. The data analysis procedure has been described by \citet{Cernicharo2022}. 
The main beam efficiency
measured during our observations varies from 0.66 at 32.4 GHz
to 0.50 at 48.4 GHz \citep{Tercero2021} and can be given across
the Q-Band by $B_\mathrm{eff}=0.797 \exp[-(\nu$(GHz)/71.1)$^2$]. The forward
telescope efficiency is 0.97. The telescope
beam size at half power intensity (hereafter referred as HPBW)
is 54.4$''$ at 32.4 GHz and 36.4$''$ at 48.4 GHz.

The intensity scale utilized in this study is the antenna temperature 
corrected for atmospheric and telescope losses ($T_A^*$).
The absolute calibration uncertainty is 10\%. However,
the relative calibration between lines within the QUIJOTE survey is certainly 
better as all of them are observed simultaneously.
All the spectral data were analysed with the GILDAS 
package\footnote{http://www.iram.fr/IRAMFR/GILDAS}.

The observed lines in the QUIJOTE band are shown in Fig. \ref{fig:TMC-1_C5H6}. Quantum numbers are indicated in each panel. In most cases the {\it ortho} and {\it para} lines are blended exactly at the same frequency producing strong features (see the bottom-right panel).
As cyclopentadiene is a nearly-oblate symmetric top, many of its {\it ortho} and {\it para} levels have close internal energies, implying nearly equal transition frequencies (see the electronic Supplementary Information for more details).
We have adopted a kinetic temperature of 9\,K \citep{agundez2023} versus the 10 K we adopted in the discovery paper \citep{cernicharo2021a}. The rest of the source parameters are the same, \ie, a
line width of 0.6 kms$^{-1}$, and a source of uniform brightness temperature with a
diameter of 80$''$ \citep[see, \eg,][]{Cernicharo2023}. The LVG calculations adopted a column density of 5 and 3.9 $\times$10$^{12}$ cm$^{-2}$ for the {\it ortho} and {\it para} species. Here, we have assumed an
{\it ortho}/{\it para} abundance ratio of 9/7. The volume density is 2$\times$10$^4$ \pcmc. The results are indicated by the red synthetic
spectra in Fig. \ref{fig:TMC-1_C5H6}. For the LTE calculations, we adopted a value of
9 K for the rotational temperature. The resulting synthetic spectra are shown in
blue in Fig. \ref{fig:TMC-1_C5H6}. As expected, no significant differences are seen between LVG
and LTE calculations. The new total column density for \ch{C5H6} of 9.9$\times$10$^{12}$ cm$^{-2}$ is slightly smaller than the previous
determination due to the different adopted kinetic temperature for the cloud.
Hence, cyclopentadiene has one of the largest column densities among the PAHs detected in TMC-1.

\section{\label{sec:concl} Conclusions}

\begin{enumerate}
   \item We present a new set of state-to-state thermal rate coefficients up to $50$~K that are calculated for the rotational (de)excitation of cyclopentadiene due to collisions with helium. Both the {\it ortho} and {\it para} nuclear spin species of \ch{C5H6} have been studied, including all rotational levels below $100$ \cmmo, \ie~a total of 222 levels per symmetry that covers rotational states from $j=0$ up to $j \leq 25$.

   \item The scattering calculations were carried out using the numerically exact close coupling (CC) and the approximate coupled states (CS) quantum scattering theories. All calculations are based on a high-level 3D potential energy surface that was developed by our group for the \ch{C5H6 - He} system, calculated from the explicitly correlated CCSD(T)-F12b {\it ab initio} method used along with an aug-cc-pVTZ basis set.

   \item Pronounced propensity trends are observed that favour the $k_c$-conserving $(\Delta k_c = 0)$ transitions. The rate coefficients exhibit some temperature dependence, which is less significant in the case of such high-magnitude transitions. The stronger dependence observed in other processes is most likely due to the high density of states, which opens many channels for the depopulation of individual levels at higher temperatures.

   \item To the best of our knowledge, this is the first study, where rotational excitation rate coefficients, calculated from the numerically exact close coupling scattering theory, have been used for non-LTE radiative transfer simulation of a large cyclic species that have been detected in the ISM. These simulations show that the explicit inclusion of state-to-state collisional rate coefficients for \ch{C5H6} has a negligible impact on the radiation (antenna) temperatures under physical conditions that are typical for cold clouds, meaning that most of the rotational levels of cyclopentadiene are fully thermalized at temperatures $\sim 10$~K and gas densities above 10$^3$ \pcmc.

   \item No significant LTE departures have been derived from our LVG radiative transfer simulations. These calculations show that the previous column density derived by \citet{Cernicharo2021b} for \ch{C5H6} is fully consistent with the LTE approximation. Nevertheless, the new complete set of rate coefficients enables a proper interpretation of upcoming potential detections in less dense astronomical environments, allowing precise non-LTE modelling when needed.

   \item A new estimation of the column density of \ch{C5H6} has been derived from the last QUIJOTE data. It is fully compatible with the previous value derived in the discovery work of cyclopentadiene \citep{Cernicharo2021b}.

   \item We have estimated the effect of collisional excitation for a molecule of similar structure to that of \ch{C5H6}, but having a dipole moment 10 times larger (a cyano derivative of cyclopentadiene for example). The density required to thermalise cyclopentadiene, $\sim$10$^3$ cm$^{-3}$, have to be multiplied, in this case, by a factor $\sim$100.

\end{enumerate}

\section*{Data availability}

The collisional data presented in this work are available in electronic supplementary material and will also be made accessible through the \href{https://emaa.osug.fr/}{EMAA}  \citep{Faure_2025}, \href{https://basecol.vamdc.eu/}{BASECOL} \citep{Dubernet_2024} and \href{https://home.strw.leidenuniv.nl/~moldata/}{LAMDA} \citep{vanderTak_2020} databases.\\

\begin{acknowledgements}
We acknowledge financial support from the European Research Council (Consolidator Grant COLLEXISM, Grant Agreement No. 811363) and the Programme National ‘Physique et Chimie du Milieu Interstellaire’ (PCMI) of CNRS/INSU with INC/INP cofunded by CEA and CNES. We wish to acknowledge the support from the CEA/GENCI for awarding us access to the TGCC/IRENE High Performance Computers within the A0150413001 project, and also the Digital Government Development and Project Management Ltd. for awarding us access to the Komondor HPC facility based in Hungary.
This article is based upon collaborations supported by the COST Action CA21101–Confined Molecular Systems: From a New Generation of Materials to the Stars (COSY), supported by COST (European Cooperation in Science and Technology).
S.D. acknowledges funding support from the Horizon Europe Marie Skłodowska-Curie Actions programme under the Grant Agreement No. 101244231 (VIBREAC). F.L. acknowledges the Institut Universitaire de France.
The authors are very grateful to C.T. Bop and M. Ben Khalifa for their contributions and essential discussions.
J. Cernicharo and M. Agúndez thank ERC for funding support under grant through grant ERC-2013-Syg-610256-NANOCOSMOS. 
They also thank Spanish MICIU (AEI/10.13039/501100011033) for funding support through projects
PID2019-106110GB-I00, 
PID2019-106235GB-I00,
PID2023-147545NB-I00, and
PID2022-137980NB-I00.
\end{acknowledgements}
%
\bibliographystyle{aa} 
\bibliography{references.bib}
%
\end{document}